\documentclass[preprint,aps]{revtex4}

\newcommand{\be}{\begin{equation}} \newcommand{\ee}{\end{equation}} 
\newcommand{\bea}{\begin{eqnarray}}\newcommand{\eea}{\end{eqnarray}}

\begin{document}
\preprint{
quant-ph/0501087}
\title{Exactly solvable non-Hermitian Jaynes-Cummings-type Hamiltonian
admitting entirely real spectra from supersymmetry}
\author{ Pijush K. Ghosh} \email{pijush@theory.saha.ernet.in}
\affiliation{Theory Division, Saha Institute of Nuclear Physics,\\ 
Kolkata 700 064, India.}
\begin{abstract} 
It is shown that for a given hermitian Hamiltonian possessing supersymmetry,
there is always a non-hermitian Jaynes-Cummings-type Hamiltonian(JCTH)
admitting entirely real spectra. The parent supersymmetric Hamiltonian
and the corresponding non-hermitian JCTH are simultaneously diagonalizable.
The exact eigenstates of these non-hermitian Hamiltonians are constructed
algebraically for certain shape-invariant potentials, including a non-hermitian
version of the standard Jaynes-Cummings(JC) model for which the parent
supersymmetric Hamiltonian is the superoscillator. It is also shown that a
non-hermitian version of the several physically motivated generalizations of
the JC model admits entirely real spectra. The
positive-definite metric operator in the Hilbert space is constructed
explicitly along with the introduction of a new inner product structure, so
that the eigenstates form a complete set of orthonormal vectors and the
time-evolution is unitary.
\end{abstract}
\maketitle
\newpage

\section{introduction}

One of the standard axioms of quantum mechanics is to consider self-adjoint
operators so that the corresponding eigenvalues are real and the
time-evolution of the eigenstates is unitary. However, a new viewpoint emerging
in the current literature is that although the condition of hermiticity is
sufficient to have a unitary theory with real spectra, it is not necessary.
The current development in this direction was boosted by the discovery of
a class of non-hermitian Hamiltonians, invariant under the combined parity and
time-reversal symmetry ($PT$), that admit real spectra for unbroken $PT$
symmetry\cite{bend,ali,ddt,zno,bag,psusy,nsusy,and,ex3,jap,hull,bpm,nc,kumar,
bm,field,quasi,quasi1}. 
The spectra appear in complex-conjugate pairs, if the $PT$ symmetry is broken
spontaneously. This class of non-hermitian Hamiltonians with unbroken $PT$
symmetry is also shown to have a further symmetry $C$ similar to the
charge-conjugation\cite{bend}.
The probabilistic interpretation of the quantum mechanics and the unitary time
evolution of the eigenstates can be restored with the construction of a new
inner-product using the $CPT$ symmetry. Field theoretical models with similar
features have been studied in the literature\cite{field}.

A complementary approach in constructing physically meaningful theories
with non-hermitian Hamiltonians admitting real spectra is to introduce the
notion of pseudo-hermiticity\cite{ali}.
An operator is said to be pseudo-hermitian, if it is related to its hermitian
adjoint through a similarity transformation. The non-hermitian Hamiltonians
admitting real spectra are shown to be pseudo-hermitian and are invariant under
an anti-linear symmetry which reduces to the standard $PT$ symmetry for some
cases. A new inner product structure along with a positive-definite
metric operator can be constructed in the Hilbert space using the property
of pseudo-hermiticity. The probabilistic interpretation of the quantum
mechanics and the unitary time evolution of the eigenstates can be restored
with this new inner product.

The purpose of this paper is to construct a new class of non-hermitian
Hamiltonians that admits entirely real spectra and allows an explicit
construction of the positive-definite metric operator in the Hilbert space.
In particular, we show that for a given hermitian Hamiltonian possessing
supersymmetry, there is always a non-hermitian JCTH\cite{jc,tc} admitting
entirely real spectra. The reality of
the spectra is derived solely by using the superalgebra and without any
particular representation of the supercharges. Moreover, the
parent supersymmetric Hamiltonian and the corresponding non-hermitian JCTH can
be diagonalized simultaneously. Thus, if the eigenstates of the parent
supersymmetric Hamiltonian are known exactly, the eigenstates of the
corresponding non-hermitian JCTH can be calculated easily. Among several
known exactly/partly solved supersymmetric quantum mechanical
Hamiltonians\cite{khare,ab1,piju,pauli}, the eigenstates
of non-hermitian JCTH corresponding to the one dimensional `shape invariant'
potentials are obtained exactly and algebraically. Only those `shape invariant'
potentials are considered for which the partner potentials are related to each
other through a translation in the parameter. One of the most notable among
these examples of non-hermitian JCTH is the standard JC
Hamiltonian with non-hermitian interaction for which the superoscillator is
the parent Hamiltonian.
We show that the non-hermitian JC model considered in this paper admits a
complete set of bi-orthonormal vectors\cite{bi}. Moreover, it can be shown
that the non-hermitian JCTH is in fact pseudo-hermitian as well as
quasi-hermitian\cite{quasi,quasi1}.

We also study a class of non-hermitian
$2 \times 2$ dimensional matrix Hamiltonians that admits entirely real
spectra. Special cases of this class of Hamiltonians include JC model with,
intensity dependent coupling\cite{bs}, Kerr nonlinearity\cite{kerr},
multi-photon interaction\cite{mp}, q-oscillator\cite{qosc} and dressed JC
model\cite{djc}. We have also shown that a non-hermitian version of the
Tavis-Cummings Model(TCM)\cite{tc}, an $N$-molecule generalization of JC model,
admits entirely real spectra.  We explicitly construct the positive-definite
metric operator
in the Hilbert space and introduce the associated inner product structure for
all these models. Consequently, the probabilistic interpretation of quantum
mechanics and the unitary time evolution of the eigenstates can be retained.

The plan of this paper is the following. We first describe the 
superalgebra and introduce the non-hermitian Hamiltonian that admits
entirely real spectra in the next section. We then find the expression for
the energy eigenvalues of this non-hermitian Hamiltonian in terms of the energy
eigenvalues of the parent supersymmetric Hamiltonian. In Sec. III, several
examples of non-hermitian JCTH admitting entirely real spectra are studied.
The example of a non-hermitian generalization of the standard JC
Hamiltonian is studied in detail in Sec. III.A, followed by a general discussion
on a non-hermitian $2 \times 2$ matrix Hamiltonian admitting entirely real
spectra in Sec. III.B. We introduce and study a non-hermitian version of the
TCM in Sec. III.C. A non-hermitian generalization of the
supersymmetric Hamiltonians with `shape-invariant' potentials for which
the partner potentials are related to each other through a translation in
the parameter is discussed in Sec. III.D. Finally, we conclude by summarizing
our results in Sec. IV. In appendix A, the metric operator for non-hermitian
$4 \times 4$ and $8 \times 8$ dimensional matrix Hamiltonian is constructed
explicitly.

\section{JCTH from supersymmetry: general formulation}

The algebra governing the ${\cal{N}}=1$ supersymmetric quantum mechanics with
the Hamiltonian $H$ is given by,
\be
\{Q, Q^{\dagger} \} = H, \ \ Q^2={Q^{\dagger}}^2=0, \ \
[H, Q] = [H, Q^{\dagger}]=0,
\label{eq1}
\ee
\noindent where $Q$ is the supercharge and $Q^{\dagger}$ is its adjoint.
The Hamiltonian $H$ is hermitian and semi-positive definite by construction.
All the eigenvalues $E_n$ of $H$ are thus real and semi-positive definite.
We now construct a non-hermitian Hamiltonian ${\cal{H}}$ corresponding to
$H$ as,
\be
{\cal{H}} =  \{ Q, Q^{\dagger} \} + c_1 e^{i \theta} Q +
c_2 e^{-i \theta} Q^{\dagger},
\label{eq6}
\ee
\noindent where $c_1$, $c_2$ and $\theta$ are real parameters. The Hamiltonian
${\cal{H}}$ is non-hermitian for $c_1 \neq c_2$ and hermitian for $c_1=c_2$. 
The Hamiltonian ${\cal{H}}$ is identical to the JC model if
$c_1=c_2$ and $Q$ is chosen to be that of one dimensional superoscillator.
We thus refer to the whole class of ${\cal{H}}$ with arbitrary $Q$ and $c_{1,2}$
as JCTH. We will show that ${\cal{H}}$ admits entirely real spectra for
\be
\beta \equiv c_1 c_2 > 0.
\label{eq4}
\ee
\noindent The spectra appears in complex conjugate pairs for $\beta < 0$.
No general conclusions could be drawn within our approach for the critical case
$\beta=0$, i.e. either $c_1=0$ or $c_2=0$. We will be working in this paper
only with $\beta$ obeying inequality (\ref{eq4}).

Define an operator linear in the supercharges $Q$ and $Q^{\dagger}$,
\be
S(\theta) \equiv c_1 e^{i \theta} Q + c_2 e^{-i \theta} Q^{\dagger}.
\label{eq2}
\ee
\noindent
The square of the operator $S$ is proportional to $H$,
\be
S^2 = \beta H.
\label{eq3}
\ee
\noindent Although the operator $S$ is non-hermitian for $c_1 \neq c_2$,
its square $S^2$ is hermitian for any real $c_{1,2}$ and semi-positive
definite for $\beta > 0$. For hermitian $S$, i.e. $c_1=c_2$, the
condition $\beta > 0$ is satisfied automatically. The Hamiltonian $H$ and the
operator $S$ can be diagonalized simultaneously, since they commute with each
other. From Eq. (\ref{eq3}), we find that the eigenvalues $E^s_n$ of $S$,
\be
E^s_n = \pm \sqrt{\beta E_n}
\label{eq5}
\ee
\noindent are real, since $E_n \geq 0$ and $\beta$ is taken to be positive.
For those cases for which $E_n$ has no upper bound, the operator
$S$ is not bounded from below due to its negative eigenvalues. The eigenvalues
of ${\cal{H}}= H + S$ are,
\bea
{\cal{E}}_n^{\pm} & \equiv & E_n + E^s_n\nonumber \\
& = & E_n \pm \sqrt{\beta E_n}.
\label{eq7}
\eea
Although ${\cal{H}}$ is not hermitian, we have the
remarkable result that its eigenvalues are real for $\beta > 0$.
Finally we remark that for $\beta < 0$, the eigenvalues
of ${\cal{H}}$ appear in complex conjugate pairs.

Following comments are in order.\\
\noindent (i) In supersymmetric quantum mechanics, the superalgebra with
arbitrary ${\cal{N}}$ number of supercharges reads\cite{aku},
\be
\{ Q_a, Q_b^{\dagger} \} = \delta_{ab} H, \ \
\{ Q_a, Q_b \} = \{ Q_a^{\dagger}, Q_b^{\dagger} \} = 0, \ \
[H, Q_a]=0=[H,Q_a^{\dagger}], \ \ a, b=1, 2, \dots {\cal{N}}.
\ee
\noindent Define the operator $S_{\cal{N}}$,
\be
S_{\cal{N}} = \sum_a \left ( c_1^a \ e^{i \theta_a} \ Q_a +
c_2^a \ e^{-i \theta_a} \ Q_a^{\dagger} \right ),
\ee
\noindent where $c^a_{1,2}$ and $\theta_a$ are real parameters for all $a$.
We now find that $S^2_{\cal{N}} = \beta_{\cal{N}} H$ with
$\beta_{\cal{N}}= \sum_a c^a_1 c^a_2$. Thus, the Hamiltonian
${\cal{H}}_{\cal{N}}=H + S_{\cal{N}}$ admits entirely real
spectra for $\beta_{\cal{N}} > 0$. The Hamiltonian ${\cal{H}}_N$ and $H$
can be diagonalized simultaneously, since $[H, S_{\cal{N}}]=0$.

\noindent (ii) There are many square-roots of the semi-positive definite
operator $S^2$. First note that for any given non-hermitian square-root,
its hermitian adjoint is also a square-root, since $S^2$ is hermitian.
A few of these non-hermitian square-roots apart from $S$ are, 
\bea
&&S_1 (\theta) = i \gamma_5 S(\theta),\nonumber \\
&& S_2 (\theta) = c_1 e^{i \theta} \ (H + \epsilon^2) \ Q + c_2 e^{-i \theta}
(H + \epsilon^2)^{-1} \ Q^{\dagger}, \ \
\{\epsilon \in \Re | \epsilon \neq 0 \},\nonumber \\
&& S_3 = i \gamma_5 S_2,
\eea
\noindent where $\gamma_5^2=1$ and it anticommutes with both $Q$ and
$Q^{\dagger}$. A construction of $\gamma_5$ is given in appendix-A.
The Hamiltonian ${\cal{H}}$ with $S$  replaced by any one of these $S_i$'s,
i.e. ${\cal{H}}_i = H + S_i$, would admit entirely real spectra for $i=1, 2, 3$.
The Hamiltonian ${\cal{H}}$ and $ {\cal{H}}_i$ are isospectral
Hamiltonians. The eigenstates of ${\cal{H}}_i$ and
$H$ can also be simultaneously diagonalized, since $[H, S_i]=0 \ \forall \ i$.
However, a separate analysis is needed in each case for constructing
the simultaneous eigenstates in the defining Hilbert space.

\noindent (iii) The relation between the JC model and supersymmetry was first
noted in \cite{sjc}, where $c_1=c_2^{\dagger}$ was treated as an
anticommuting variable instead of a $c$-number. This anticommuting variable
also anticommutes with all the odd operators of the superalgebra and
$c_1^{\dagger} c_1$ was chosen to be nilpotent in order to  derive the
spectrum. Thus, the approach taken in this paper in relating JC model to
supersymmetry is different from that of Ref. \cite{sjc}. Our approach is
similar to that of Ref. \cite{ab2} and the Hamiltonian ${\cal{H}}$ should be
treated as a matrix Hamiltonian rather than an element of the superalgebra.
However, the Hamiltonian $H$ is indeed a bosonic element of the superalgebra
and the underlying superalgebra greatly simplifies the whole analysis.

\section{ Examples of Non-hermitian JCTH}

The superalgebra (\ref{eq1}) for one dimensional quantum mechanical system
can be realized in terms of the Pauli matrices $\sigma_i$'s as,
\be
Q = \sigma_+ a, \ \ Q^{\dagger} = \sigma_- a^{\dagger}, \ \ \
\sigma_{\pm} = \frac{1}{2} \left ( \sigma_1 \pm
i \sigma_2 \right ),
\label{eq8}
\ee
\noindent where the operators $a$ and $a^{\dagger}$ are functions of the
position and the momentum only. The Hamiltonian ${\cal{H}}$ now reads,
\be
{\cal{H}} = a^{\dagger} a + \frac{1}{2} [a, a^{\dagger}] \left (\sigma_3 +
1 \right ) + c_1 e^{i \theta} \ \sigma_+ a + c_2 e^{-i \theta} \
\sigma_- a^{\dagger}.
\label{eq9}
\ee
\noindent The Hamiltonian ${\cal{H}}$ and its adjoint ${\cal{H}}^{\dagger}$
are related to each other through the transformation $c_1 \leftrightarrow c_2$.
With the introduction of an operator $\eta$,
\be
\eta \equiv \left ( \matrix{{\delta_1} & {0}\cr \\
{0} & {\delta_2}} \right ), \ \ \ \frac{\delta_1}{\delta_2} = {\frac{c_2}{c_1}},
\label{et}
\ee
\noindent it is easily verified that ${\cal{H}}$ is pseudo-hermitian,
i. e. ${\cal{H}}^{\dagger} = \eta \ {\cal{H}} \ \eta^{-1}$. It may be noted
that $\eta$ reduces to an identity matrix multiplied by a constant 
$\delta_1=\delta_2$ in the limit $c_1=c_2$ for which the hermiticity of
${\cal{H}}$ is restored. The matrix $\eta$ is not unique and among all possible
such matrices, the unique positive-definite matrix $\eta_+$ is obtained by
taking $\delta_1=\gamma^{-1}$ and $\delta_2=\gamma$, where $\gamma$ is defined
as $\gamma \equiv \sqrt{\frac{c_1}{c_2}}$. Note that the metric $\eta_+$ is
unique up to an overall multiplication factor that is positive-definite.
The condition of pseudo-hermiticity implies that\cite{ali}, if
$| \psi \rangle$ is an eigenstate of ${\cal{H}}$ with the eigenvalue
${\cal{E}}$, then,
\be
|\phi \rangle = \eta \ |\psi \rangle,
\label{rel}
\ee
\noindent is an eigenstate of ${\cal{H}}^{\dagger}$ with the same eigenvalue
${\cal{E}}$. Conversely, if $|\tilde{\phi} \rangle$ is an eigenstate of
${\cal{H}}^{\dagger}$ with the eigenvalue $\tilde{\cal{E}}$, then
$|\tilde{\psi} \rangle = \eta^{-1} |\tilde{\phi} \rangle $ is an eigenstate of
${\cal{H}}$ with the eigenvalue $\tilde{\cal{E}}$. The relation (\ref{rel})
and the positive-definite matrix $\eta_+$ will be used later to construct a
new inner product in the Hilbert space.
Further, ${\cal{H}}$ can be mapped to a hermitian Hamiltonian
$h$,
\be
h = a^{\dagger} a + \frac{1}{2} [a, a^{\dagger}] \left (\sigma_3 +
1 \right ) + \beta \ \left ( e^{i \theta} \sigma_+ a + e^{-i \theta}
\ \sigma_- a^{\dagger} \right ),
\ee
\noindent through the similarity transformation,
$h = \rho {\cal{H}} \rho^{-1}$, where $\rho := \sqrt{\eta}$ is the positive
square-root of $\eta$. In the notion of Ref. \cite{quasi,quasi1}, ${\cal{H}}$
is also quasi-hermitian.

\subsection{ Non-hermitian resonant JC model}

The Hamiltonian ${\cal{H}}$ reduces to a non-hermitian generalization of the
standard JC model, if $a,a^{\dagger}$ are chosen to represent the
usual anhilation, creation operators of harmonic oscillators, i.e.
$a=\frac{1}{\sqrt{2}} ( p - i x),
a^{\dagger} =\frac{1}{\sqrt{2}} ( p+ i x), [a, a^{\dagger}]=1$.
The ground state has zero energy with the eigenstate,
$ | \psi_{0} \rangle = \gamma^{-\frac{1}{2}}
\left [ \matrix{ {0 }\cr \\ { |0 \rangle}} \right ]$,
where $|n \rangle$ is the standard orthonormal basis of the harmonic
oscillator with $a | n \rangle = \sqrt{n} |n-1 \rangle$, 
$ a^{\dagger} | n \rangle = \sqrt{n+1} |n+1 \rangle$,
$|n \rangle = \frac{(a^{\dagger})^n}{\sqrt{n}} |0 \rangle$ and $|0\rangle$ is
the vacuum state annihilated by $a$. The excited states are,
\be 
{\cal{E}}_{n+1}^{\pm} = n + 1 \pm \sqrt{\beta (n+1)},\ \ \
| \psi_{n+1}^{\pm} \rangle = (2 \gamma)^{-\frac{1}{2}}
\left [ \matrix{ {\pm e^{i \theta} \gamma |n \rangle}\cr \\
{ |n+1 \rangle}}
\right ], \ \ n=0, 1, 2, \dots
\label{eq10}
\ee
\noindent The eigenvalues of the adjoint Hamiltonian ${\cal{H}}^{\dagger}$ are
still given by ${\cal{E}}_{n+1}^{\pm}$, while the corresponding eigenvectors
$|\phi_{n+1}^{\pm} \rangle$ are obtained from $|\psi_{n+1}^{\pm} \rangle$
through the transformation $c_1 \leftrightarrow c_2$ or using the Eq.
(\ref{rel}) with $\eta_+$. In particular,
\be
| \phi_{n+1}^{\pm} \rangle = (2 \gamma)^{-\frac{1}{2}}
\left [ \matrix{ {\pm e^{i \theta} |n \rangle}\cr \\ {\gamma |n+1 \rangle}}
\right ], \ \ n=0, 1, 2, \dots
\ee
\noindent Note that the state $|\psi_n^{\pm} \rangle$ is not orthogonal to
$|\psi_m^{\mp} \rangle$ with the usual definition of inner product, unless
the condition of hermiticity, i.e., $c_1=c_2$ is imposed. However, the
eigenstates $| \psi_{n}^{\pm} \rangle$ and $|\phi_{m}^{\pm} \rangle$ together
constitute a complete set of bi-orthonormal vectors for arbitrary $c_{1,2}$,
\be
\langle \psi_n^{I} | \phi_m^{J} \rangle = \delta_{mn} \delta_{IJ}, \ \
\sum_{n,I} |\psi_n^I \rangle \langle \phi_n^I |=
\sum_{n,I} |\phi_n^I \rangle \langle \psi_n^I |= {\bf 1}, \ \ I, J= +, - .
\ee
\noindent A new inner product structure can be constructed as,
$\langle \langle u | v \rangle \rangle_{\eta_+} =
\langle u | \eta_+ | v \rangle$.
The norm of any arbitrary state vector $\langle u | = (\langle m|, \langle n|)$ is positive definite under this new inner product structure, since $\gamma$ is
always positive definite for $\beta > 0$. Furthermore, note that $\eta_+$
reduces to an identity matrix in the limit $c_1=c_2$ for which the hermiticity
of ${\cal{H}}$ is restored. Consequently, the new inner product structure 
is identical to the standard one for $c_1=c_2$. We now find a complete set
of orthonormal vectors for ${\cal{H}}$,
\be
\langle \langle \psi_{n}^{\pm} | \psi_{m}^{\pm} \rangle \rangle_{\eta_+}
=\delta_{nm}, \ \
\langle \langle \psi_{n}^{\pm} | \psi_{m}^{\mp} \rangle \rangle_{\eta_+} =0, \ \
\sum_{m} \eta_+ \left ( |\psi_m^{-} \rangle \langle
\psi_m^- | + |\psi_m^{+} \rangle \langle \psi_m^+ | \right ) ={\bf 1}.
\ee
\noindent The Hamiltonian ${\cal{H}}$ is hermitian with respect to this new
inner product.

A comment is in order at this point. The operator $\gamma_5$ in Eq. (\ref{gaga})
is determined as $\gamma_5=- \sigma_3$ for $N=1$ and $\psi_1 :=\sigma_-$. 
Note that $S_1(\theta+\frac{\pi}{2})=S(\theta)$, implying that ${\cal{H}}_1$
and ${\cal{H}}$ are not independent of each other. Similarly,
${\cal{H}}_2$ and ${\cal{H}}_3$ are also not independent. However, it is
expected that these Hamiltonians will be independent of each other for $N > 1$. 

\subsection{Non-hermitian non-resonant JC model \& other generalizations}

The example considered in the previous section corresponds to a non-hermitian
generalization of the resonant JC model. A non-hermitian version of the
non-resonant JC model also admits entirely real spectra. In particular,
\be
{\cal{H}}^{NR} = {\cal{H}} + \Delta \sigma_3,
\ee
\noindent is pseudo-hermitian under $\eta$ and admits entirely real spectra
for $\beta > 0$. For $\beta < 0$, the eigenvalues are not entirely real and 
a further choice of $\theta=0$ reproduces the result obtained in
Ref. \cite{bpm}. The eigenvalues are,
\be
{\cal{E}}^{\pm}_{n+1} = n + 1 \pm \left [ \Delta^2 +
\beta (n+1) \right ]^{\frac{1}{2}},
\ee
\noindent with the eigenstates,
\be
|\psi_{n+1}^{\pm} \rangle = \left ( \frac{\gamma}{\gamma^2 +
{\mid \Gamma_{\pm}^n \mid}^2}
\right )^{\frac{1}{2}} \left [ \matrix{ {\Gamma_{\pm}^n |n \rangle}\cr \\
{|n+1 \rangle}} \right ], \ \
\Gamma_{\pm}^n \equiv \frac{e^{i \theta}}{c_2 \sqrt{n+1}}
\left [ \Delta \pm \left
\{\Delta^2+ \beta (n+1) \right \}^{\frac{1}{2}} \right ],
\ee
\noindent where $n$ is a non-negative integer. The ground-state energy is,
${\cal{E}}_0= - \Delta$ with the corresponding eigenstates,
$|\psi_{0} \rangle = \gamma^{-\frac{1}{2}} \left [ \matrix{ {0 }\cr \\
{ |0 \rangle}} \right ]$.
These eigenstates form a complete set of orthonormal vectors
with respect to the new inner product structure $\langle \langle . | .
\rangle \rangle _{\eta_+}$.

There are many physically motivated generalizations of the standard JC model by
including intensity-dependent coupling\cite{bs}, Kerr nonlinearity\cite{kerr},
multi-photon interaction\cite{mp} and q-oscillator interaction\cite{qosc}. In
order to study a non-hermitian version of these generalized models, consider the
following non-hermitian matrix Hamiltonian,
\be
\tilde{\cal{H}} = \left ( \matrix{{f_1(a, a^{\dagger})} &
{c_1 e^{i \theta} \ g(a, a^{\dagger})}\cr\\
{c_2 e^{-i \theta} \ g^{\dagger}(a, a^{\dagger})} & 
{f_2(a, a^{\dagger})}} \right ), \ \ f_1^{\dagger} = f_1, \ \
f_2^{\dagger} = f_2,
\ee
\noindent where the hermitian adjoint is taken with respect to the standard
inner product. The Hamiltonian $\tilde{\cal{H}}$ may or may not be
supersymmetric
for $c_1=c_2=0$. In general, $f_1, f_2$ and $g$ are arbitrary functions of the
operators $a, a^{\dagger}$. Specific choices of $f_1, f_2$ and $g$ lead to
integrable reductions of $\tilde{\cal{H}}$ that include a non-hermitian
version of the generalized JC models\cite{bs,kerr,mp,qosc}.
Note that the Hamiltonian $\tilde{\cal{H}}$ in its full generality is
pseudo-hermitian, $\tilde{\cal{H}}^{\dagger} = \eta \tilde{\cal{H}}
\eta^{-1}$, where $\eta$ is given by Eq. (\ref{et}). Following Ref. \cite{ali},
we conclude that $\tilde{\cal{H}}$ admits entirely real spectra for
$\beta > 0$, since the positive definite metric $\eta_+$ exists for a positive
definite $\beta$.
Further, the new inner product structure $\langle \langle . | .
\rangle \rangle _{\eta_+}$ should be used for all relevant calculations.
The Hamiltonian $\tilde{\cal{H}}$ is also quasi-hermitian,
$\tilde{h} = \rho \tilde{\cal{H}} \rho^{-1}$, where $\tilde{h}$ is hermitian,
\be
\tilde{h} = \left ( \matrix{{f_1(a, a^{\dagger})} &
{\beta \ e^{i \theta} g(a, a^{\dagger})}\cr\\
{\beta \ e^{-i \theta} g^{\dagger}(a, a^{\dagger})} & 
{f_2(a, a^{\dagger})}} \right ),
\ee
\noindent and $\rho:=\sqrt{\eta}$. The eigen-spectra of $\tilde{\cal{H}}$
corresponding to the non-hermitian generalization of the models considered
in Ref. \cite{bs,kerr,mp,qosc} could be obtained in a straightforward way.

The customary JC model is derived adopting the rotating-wave
approximation\cite{jc}. There are additional terms in the Hamiltonian
without this approximation and the corresponding Hamiltonian is known as
``dressed" JC  Hamiltonian in the literature\cite{sjc,djc}. Note that a
non-hermitian version of the ``dressed" JC model is included in the
Hamiltonian $\tilde{\cal{H}}$. In particular, if we choose,
\be
f_1 = b_1 a^{\dagger} a + b_2 ( a^2 + {a^{\dagger}}^2 ) + b_3, \ \
f_2 = d_1 a^{\dagger} a + d_2 ( a^2 + {a^{\dagger}}^2 ) + d_3, \ \
g= e_1 a + e_2 a^{\dagger},
\ee
\noindent where $b_i, d_i, e_i$ are arbitrary real parameters, then
${\tilde{\cal{H}}}$ is a non-hermitian version of the ``dressed" JC  model
that admits real spectra. All the discussions in the previous paragraph
holds true for this Hamiltonian, except for the fact that finding the exact
spectra is a non-trivial task. Finally, we would like to mention
that for the special choices of the parameters,
\be
b_1=d_1=e_1^2+e_2^2, \ \
b_2=d_2=e_1 e_2, \ \ b_3= e_1^2, \ \ d_3=e_2^2,
\ee
\noindent the diagonal elements $f_1$ and $f_2$ correspond to the
partner Hamiltonians of a supersymmetric Hamiltonian $\tilde{H}=
\{\tilde{Q}, \tilde{Q}^{\dagger} \}$ with $\tilde{Q}=\sigma_+ \ g,
\tilde{Q}^{\dagger} = \sigma_- g^{\dagger}$. The Hamiltonian $\tilde{\cal{H}}$
can be expressed as, $\tilde{\cal{H}}= \tilde{H} + c_1 e^{i \theta} \tilde{Q} +
c_2 e^{- i \theta} \tilde{Q}^{\dagger}$.

\subsection{Non-hermitian Tavis-Cummings model}

The TCM deals with $N$ identical two-level molecules
interacting through a dipole coupling with a single-mode quantized radiation
field\cite{tc}. We consider a non-hermitian version of TCM,
\be
H_{TC}=a^{\dagger} a + R_3 + \frac{1}{2} + c_1 e^{i \theta} a R_+ +
c_2 e^{-i \theta} a^{\dagger} R_-,
\ee
\noindent where the generators $R_3, R_{\pm}$ satisfy the $SU(2)$ algebra,
\be
[R_3, R_{\pm} ] = \pm R_{\pm}, \ \
[R_+, R_-] = 2 R_3.
\label{su2}
\ee
\noindent The hermitian TCM\cite{tc} is obtained in the limit $c_1=c_2$.
Without loss of any generality, we have added an extra term equal to
$\frac{1}{2}$ in the expression of $H_{TC}$ for the convenience of discussions
in the later part of this section. The $SU(2)$ generators are realized in terms
of the Pauli matrices and the $ 2 \times 2$ identity matrix $I$ as,
\bea
&& R_{\mp} = \sum_{i=1}^N \Sigma^i_{\mp}, \ \
R_3 = \sum_{i=1}^N \Sigma^i_3,\nonumber \\
&& \Sigma^i_{a} = I \otimes \dots \otimes I \otimes \sigma_{a} \otimes I
\otimes \dots \otimes I, \ \ a= \mp, 3 \ \ (\sigma_a \ in \ i-th \ position ).
\label{pauli}
\eea
We now define the metric operator $\eta_N$ as,
\be
\eta_N = \eta \otimes \eta \otimes \dots \otimes \eta \ \ ( N \ times).
\label{metric}
\ee
\noindent The Hamiltonian $H_{TC}$ is pseudo-hermitian under $\eta_N$,
$ H^{\dagger}_{TC} = \eta_N \ H_{TC} \ \eta^{-1}_N$. The positive-definite
metric operator $\eta_N^+$ exists and it can be constructed through the
replacement of $\eta$ by $\eta_+$ in $\eta_N$. Thus, the non-hermitian
TCM admits entirely real spectra with consistent quantum mechanical
interpretation for $\beta > 0$. The Hamiltonian $H_{TC}$ is quasi-hermitian
under $\rho:= \sqrt{\eta_N}$, $ h_{TC} = \rho \ H_{TC} \ \rho^{-1}$, where
the hermitian $ h_{TC} = a^{\dagger} a + R_3 + \frac{1}{2}
+ \beta ( e^{i \theta} a R_+ + e^{-i \theta} a^{\dagger} R_-)$.

The individual two-level molecules in the TCM, with the $SU(2)$ generators
given by Eq. (\ref{pauli}), are independent of each other. However, the
individual molecules are no longer independent of each other, if the
representation of the $SU(2)$ generators given by Eq. (\ref{gen}) in
Appendix-A is used. The advantage of the representation of the $SU(2)$
generators given by Eq. (\ref{gen}) is that the Hamiltonian $H_{TC}$ can
now be written as,
\be
H_{TC} = \{ Q, Q^{\dagger} \} + c_1 e^{i \theta} Q +
c_2 e^{-i \theta} Q^{\dagger}, \ \
Q=a^{\dagger} R_-, \ \ Q^{\dagger} = a \ R_+.
\ee
\noindent Defining the metric operator and its inverse as,
\be
\eta = \delta_2 R_+ R_- + \delta_1 R_- R_+, \ \
\eta^{-1} = \delta^{-1}_2 R_+ R_- + \delta^{-1}_1 R_- R_+,
\ee
\noindent we find that $H^{\dagger}_{TC} = \eta H_{TC} \eta^{-1}$.
The positive-definite metric exists and can be obtained by taking
$\delta_1=\gamma^{-1}, \delta_2=\gamma$. Thus, the TCM with the
representation of the $SU(2)$ generators given by Eq. (\ref{gen})
also admits a real spectra with consistent quantum mechanical interpretation.

\subsection{Non-hermitian JCTH \& Shape-invariant potentials}

The operator $a, a^{\dagger}$ corresponding to  more general supersymmetric
Hamiltonian $H$ can be written as,
\be
a= p - i w(x,q), \ \ a^{\dagger} = p + i w(x,q),
\label{eq11}
\ee
\noindent where $q$ is a parameter and the superpotential $w$ is a real
function of the coordinate $x$. The supersymmetric Hamiltonian $H$ and the
corresponding non-hermitian JCTH ${\cal{H}}$ now reads,
\bea
&& H = p^2 + w^2 + i \sigma_3 [p, w], \ \
{\cal{H}} = \Pi^2 + W^2 + i \sigma_3 \left [ \Pi, W \right ],
\eea
\noindent where the operators $\Pi$ and $W$ are defined as,
\be
\Pi \equiv p + \frac{c_1 e^{i \theta} }{2} \sigma_+ +
\frac{c_2 e^{-i \theta}}{2} \sigma_-,\ \ \
W \equiv w(x,q) + \frac{c_1}{2} e^{i (\theta-\frac{\pi}{2}) } \sigma_+ +
\frac{c_2}{2} e^{-i (\theta-\frac{\pi}{2})} \sigma_-.
\ee
\noindent Both $\Pi$ and $W$ are non-hermitian for $c_1 \neq c_2$ and
$i \sigma_3 \left [ \Pi, W \right ] = i \sigma_3 [p, w] - \frac{c_1 c_2}{2}$.
The supersymmetric Hamiltonian $H$ is diagonal with the diagonal elements, 
$H_{\pm} \equiv p^2 + V_{\pm}$, where the partner potentials $V_{\pm}$
are defined as, $V_{\pm} \equiv w^2 \pm \frac{\partial w}{\partial x}$.

We consider here only those cases for which $V_{\pm}$ are related to each other
through a translation in the parameter $q$. Examples of such `shape-invariant'
potentials are abundant in the literature\cite{khare}. Following
Ref. \cite{ab1,ab2}, we further introduce the operators,
\be
T \equiv exp \left ( \xi \frac{\partial}{\partial q} \right ), \  \
T^{\dagger} \equiv exp\left ( - \xi \frac{\partial}{\partial q}\right ), \ \
B _- \equiv T^{\dagger}(q) a(q), \ \ B_+ \equiv a^{\dagger}(q) T(q), 
\label{eq12}
\ee
\noindent where $T$ is the translational operator for the parameter $q$.
In particular, $T$ acting on an operator $O$ gives 
$ T(q) O(q) T^{-1}(q) = O(q+\xi)$.
Generalizing the algebraic method employed in \cite{ab2} to the non-hermitian
case, we find the eigenstates of ${\cal{H}}$,
\be 
{\cal{E}}_{n+1}^{\pm} = E_{n+1} \pm \sqrt{\beta E_{n+1}}, \ \
| \psi_{n+1}^{\pm} \rangle = (2 \gamma )^{-\frac{1}{2}}
\left [ \matrix{ {e^{i \theta} \gamma T |n \rangle}\cr \\ \pm { |n+1 \rangle}}
\right ], \ \ n=0, 1, 2, \dots,
\label{eq13}
\ee
\noindent where $|n\rangle$ is an orthonormal basis for the generalized
Heisenberg algebra satisfied by the operators $B_{\pm}$ with the
role of number operator being played by $B_+ B_-$. $E_n$ is eigenvalue of
the operator $B_+ B_-= a^{\dagger} a$. The ground state has zero energy with
the eigenstate, $ | \psi_{0} \rangle = \gamma^{-\frac{1}{2}} \left
[ \matrix{ {0 }\cr \\
{ |0 \rangle}} \right ]$. The eigenstates in Eq.(\ref{eq13}) form a complete
set of orthonormal vectors with the respect to the new inner product structure
defined above.

\section{Summary \& Conclusions}

We have shown that for a given hermitian Hamiltonian possessing supersymmetry,
there always exists a non-hermitian JCTH admitting entirely real spectra.
Moreover, if the parent supersymmetric Hamiltonian is exactly solvable, the
corresponding JCTH can also be solved exactly. This is because these two
Hamiltonians are simultaneously diagonalizable. These results are derived
solely by using the superalgebra and without any particular representation
for the supercharges. Thus, our prescription to construct non-hermitian
Hamiltonian admitting entirely real spectra is very general and constitutes a
new class itself.

We have also studied a class of non-hermitian
$2 \times 2$ dimensional matrix Hamiltonians that admits entirely real
spectra. Special cases of this class of Hamiltonians include JC model with,
intensity dependent coupling, Kerr nonlinearity, multi-photon interaction,
q-oscillator and dressed JC model. We have also shown that a non-hermitian
version of the TCM, an $N$-molecule generalization of JC model, admits entirely
real spectra.

We have solved exactly the JCTH corresponding to one dimensional supersymmetric
and shape invariant potentials. We have constructed a new inner product
structure along with the positive-definite metric operator for this class of
JCTH as well as for all other physically motivated generalizations of the JC
model considered in this paper. Thus, for all these non-hermitian models,
a complete set of orthonormal vectors exists and a consistent
interpretation of the relevant physical observables are possible.
Finally, the examples considered in this paper include a non-hermitian
generalization of the standard JC model for which the parent
supersymmetric Hamiltonian is the superoscillator. It would be nice if
any observable effect due to the non-hermiticity could be noticed for
this Hamiltonian using micromaser\cite{maser} experiments. 

\acknowledgments{This work is supported (DO No. SR/FTP/PS-06/2001) by SERC,
DST, Govt. of India through the Fast Track Scheme for Young
Scientists:2001-2002.}

\appendix{
\section{Explicit construction of the metric operator $\eta$ for a
many-particle system}
Consider a set of fermionic variables $\psi_i$ and $\psi_i^{\dagger}$
satisfying the Grassman algebra,
\be
\{\psi_i,\psi_j \}=\{\psi_i^{\dagger}, \psi_j^{\dagger} \}=0, \ \
\{ \psi_i, \psi_j^{\dagger} \} = \delta_{ij}.
\ee
\noindent One can further define an operator $\gamma_5$,
\be
\gamma_5 = (-1)^N \prod_{i=1}^N \left ( 2 \psi_i^{\dagger} \psi_i -1 \right ),
\label{gaga}
\ee
\noindent which has the following property,
\be
\gamma_5^2 = 1, \ \ \{\gamma_5, \psi_i \}= 0 = \{ \gamma_5, \psi_i^{\dagger} \}
\ \ \forall \ i.
\label{a5}
\ee
\noindent The matrix representation\cite{cliff} of these anticommuting
variables can be used to construct the generators of $SU(2)$. In particular,
\be
R_-= \frac{1}{\sqrt{N}} \sum_i \psi_i, \ \
R_+= \frac{1}{\sqrt{N}} \sum_i \psi_i^{\dagger}, \ \
R_3= \frac{1}{2 N} \sum_{i,j} \left [ \psi_i^{\dagger}, \psi_j \right ].
\label{gen}
\ee
\noindent Apart from satisfying the $SU(2)$ algebra (\ref{su2}),
the generators (\ref{gen}) also satisfy the following relations,
\be
R^2_-=0=R^2_+, \ \ \{R_-, R_+ \} = 1, \ \ R_{\pm} R_{\mp} =
\left ( \pm R_3 + \frac{1}{2} \right ).
\ee
\noindent The generators in Eq. (\ref{gen}) correspond to higher dimensional
reducible representation of $SU(2)$.

The superalgebra (\ref{eq1}) for higher dimensional or many-particle
quantum mechanical system can be realized as,
\be
Q= \sum_{i=1}^N a_i^{\dagger} \psi_i, \ \
Q^{\dagger} = \sum_{i=1}^N a_i \psi_i^{\dagger}.
\ee
\noindent The operators $a_i(a_i^{\dagger})$ are expressed in terms
of the momentum $p_i$ of the $i$th particle and the superpotential $W(x)$,
\be
a_i=p_i - i W_i(x), \ \ a_i^{\dagger} = p_i + i W_i(x), \ \
W_i(x) = \frac{\partial W}{\partial x_i},
\ee
\noindent where particle coordinates are denoted by $x_i$. Note that,
\be
[a_i, a_j ] = 0 = [a_i^{\dagger}, a_j^{\dagger} ], \ \
[a_i, a_j^{\dagger}] = \frac{\partial^2 W}{\partial x_i \partial x_j}.
\ee
\noindent The anti-commutation property of $\gamma_5$ in Eq. (\ref{a5})
implies that $\{Q, \gamma_5 \} = 0 = \{Q^{\dagger}, \gamma_5 \}$.

We now construct the metric operator $\eta$ for $N=2$
and $N=3$. For $N=2$, we choose,
\be
\psi_1 = \sigma_3 \otimes \sigma_-, \ \
\psi_2 = \sigma_- \otimes I.
\ee
\noindent The operators ${\cal{H}}, Q, S$ are $4 \times 4$ dimensional
matrices with the elements being functions of $a_{1,2}$ and $a^{\dagger}_{1,2}$.
It turns out that the metric operator $\eta$ is given by $\eta_2$ of
Eq. (\ref{metric}).
In a similar way, for $N=3$, we choose,
\be
\psi_1= \sigma_3 \otimes \sigma_- \otimes I, \ \
\psi_2= I \otimes \sigma_3 \otimes \sigma_-, \ \
\psi_3= \sigma_- \otimes I \otimes \sigma_3.
\ee
\noindent The operators ${\cal{H}}, Q, S$ are now $8 \times 8$ dimensional
matrices
with the elements being functions of $a_{1,2,3}$ and $a^{\dagger}_{1,2,3}$.
We find the metric operator $\eta$ is given by $\eta_3$ of Eq. (\ref{metric}).
Based on these results, {\it we conjecture that for arbitrary
$N$, the metric operator is given by $\eta_N$}.
The positive-definite metric operator $\eta^+_N$ is obtained through
the replacement of $\eta$ by $\eta_+$.

The Hamiltonian ${\cal{H}}$ admits entirely real spectra for $\beta > 0$.
Further, we have explicitly constructed the positive-definite metric operator
for $N=2$ and $N=3$, implying that a consistent quantum mechanics can be
constructed for these cases and for arbitrary $W(x)$. In fact, in two and three
dimensions, one can now
easily construct {\it an exactly solvable non-hermitian ${\cal{H}}$ admitting
entirely real spectra and with consistent quantum mechanical interpretation by
simply choosing an exactly solvable $H$}. Some of the simplest choices for $H$
are two and three dimensional superoscillators and all types of two
and three particle super-Calogero models.  
}

\end{document}